\documentclass[letterpaper, 10 pt, conference]{ieeeconf}  

\IEEEoverridecommandlockouts                              
\let\labelindent\relax
\usepackage[utf8]{inputenc} 
\usepackage[T1]{fontenc}    
\usepackage{times}
\usepackage{amsfonts}       
\usepackage{nicefrac}       
\usepackage{microtype}      
\usepackage{xcolor}         
\usepackage{graphicx}
\usepackage{mathtools}
\usepackage{amsmath}
\usepackage{algorithm}
\usepackage{dashrule}
\usepackage{algpseudocode}

\usepackage{subfigure}
\usepackage{float}
\usepackage{hyperref}
\usepackage{enumitem}

\newcommand{\be}{\begin{equation}}
\newcommand{\ee}{\end{equation}}
\newcommand{\Expected}[1]{\mathrm{E}\left[#1\right]}
\newcommand{\nn}{\nonumber}

\DeclareMathOperator*{\argmax}{argmax}
\newcommand{\ave}[1]{{#1}^{\mathrm{ave}}}
\newcommand{\freq}[1]{{#1}^{\mathrm{freq}}}
\newcommand{\delQ}{\delta{Q}}

\newcommand{\odelQ}{\overline{\delta Q}}

\newcommand{\TV}[1]{\left\|#1\right\|_{\mathrm{TV}}}

\newcommand{\what}[1]{\widehat{#1}}

\newcounter{theorem}
\newenvironment{lemma}[1]{\refstepcounter{theorem}\par
\noindent \textbf{Lemma \thetheorem.} \em \rmfamily}{}
\newenvironment{theorem}[1]{\refstepcounter{theorem}\par
\noindent \textbf{Theorem \thetheorem.} \em \rmfamily}{}

\title{\LARGE \bf
Episodic Logit-Q Dynamics for Efficient Learning in Stochastic Teams
}

\author{Onur Unlu and Muhammed O. Sayin
\thanks{This work was supported by the TUBITAK BIDEB 2232-B International Fellowship for Early Stage Researchers under Grant Number 121C124.}
\thanks{Onur Unlu and Muhammed O. Sayin are with the Department of Electrical \& Electronics Engineering,
        Bilkent University, Ankara, Turkey 06800. Emails:
        \tt\small{onur.unlu@ug.bilkent.edu.tr}, \tt\small{sayin@ee.bilkent.edu.tr}}%
}

\begin{document}

\maketitle
\thispagestyle{empty}
\pagestyle{empty}

\begin{abstract}

We present new learning dynamics combining (independent) log-linear learning and value iteration for stochastic games within the auxiliary stage game framework. The dynamics presented provably attain the efficient equilibrium (also known as optimal equilibrium) in identical-interest stochastic games, beyond the recent concentration of progress on provable convergence to some (possibly inefficient) equilibrium. The dynamics are also independent in the sense that agents take actions consistent with their local viewpoint to a reasonable extent rather than seeking equilibrium. These aspects can be of practical interest in the control applications of intelligent and autonomous systems. The key challenges are the convergence to an inefficient equilibrium and the non-stationarity of the environment from a single agent's viewpoint due to the adaptation of others. The log-linear update plays an important role in addressing the former. We address the latter through the play-in-episodes scheme in which the agents update their Q-function estimates only at the end of the episodes. 

\end{abstract}

\section{INTRODUCTION}

Shapley introduced stochastic games (SGs) as a generalization of Markov decision processes (MDPs) to non-cooperative multi-agent settings \cite{ref:Shapley53}. Since MDPs are the basis of reinforcement learning, SGs have also attracted significant interest as an ideal model for multi-agent reinforcement learning. Examples include planning for intelligent and autonomous systems \cite{ref:Zhang21}. Shapley also showed that Markov stationary equilibrium exists in discounted SGs with finite state and action spaces similar to the existence of a stationary solution in MDPs. Apart from the studies focusing on the computation of the stationary equilibrium in SGs, recently, there has also been a growing interest in determining whether non-equilibrium adaptation of learning agents reach stationary equilibrium in SGs, e.g., \cite{ref:Leslie20,ref:Sayin20,ref:Baudin22,ref:Sayin22}, similar to the extensive literature on learning in the repeated play of games, e.g., see \cite{ref:Fudenberg98}. However, in identical-interest SGs, also known as stochastic teams, reaching possibly inefficient equilibrium may not be desirable in control and optimization applications. For example, efficient learning in which agents reach efficient equilibrium has been studied extensively for games with repeated play, e.g., \cite{ref:Marden12,ref:Pradelski12, ref:TatianaProof}. However, very limited results exist on efficient learning in stochastic games, as reviewed in detail later. 

In the \textit{auxiliary stage game} framework, we can view SGs as agents are playing a stage game associated with the state visited. The payoffs of these stage games, called \textit{Q-functions}, depend not only on the immediate reward they receive but also on the rewards they will receive in future stages, often referred to as the continuation payoff. However, the continuation payoff depends on the evolving strategies of the agents. Hence, the stage games are not necessarily stationary. The non-stationarity of stage games is acknowledged as one of the main challenges for learning in SGs \cite{ref:Ozdaglar21}. There is no straightforward generalization of the existing results from repeated play setting that address efficient learning, e.g., \cite{ref:Marden12,ref:Pradelski12, ref:TatianaProof} into SGs. Recently, the \textit{two-timescale learning framework} addressed the non-stationarity issue for the best response and fictitious play dynamics \cite{ref:Leslie20,ref:Sayin20}. However, these learning dynamics do not necessarily reach efficient equilibrium even in identical-interest SGs.

In this paper, we present the new \textit{logit-Q} learning dynamics, combining log-linear learning and value iteration for efficient learning in SGs. We also present the \textit{independent logit-Q} learning dynamics to address the coordination burden in the synchronous (in turn) update of actions inherent to the log-linear learning. We show the almost sure convergence of the Q-function estimates to the \textit{globally optimum} Q-function of the underlying identical-interest SG in both logit-Q and independent logit-Q dynamics. We also verify the convergence of the learning dynamics via numerical examples.

Independent and simultaneous adaptation of agents poses the key challenges of: $i)$ achieving the \textit{global} optimum rather than converging to an inefficient equilibrium and $ii)$ the \textit{non-stationarity} of the environment from each agent's perspective. The former one is addressed in the log-linear learning for the repeated games yet the convergence results, such as \cite{ref:Marden12}, are in the sense that only the optimal solution is \textit{stochastically stable}. On the other hand, the latter one is viewed as a core issue in multi-agent reinforcement learning \cite{ref:Zhang21} and in learning in SGs, as discussed above.

\begin{figure*}[t!]
\centering
\includegraphics[width=.90\linewidth]{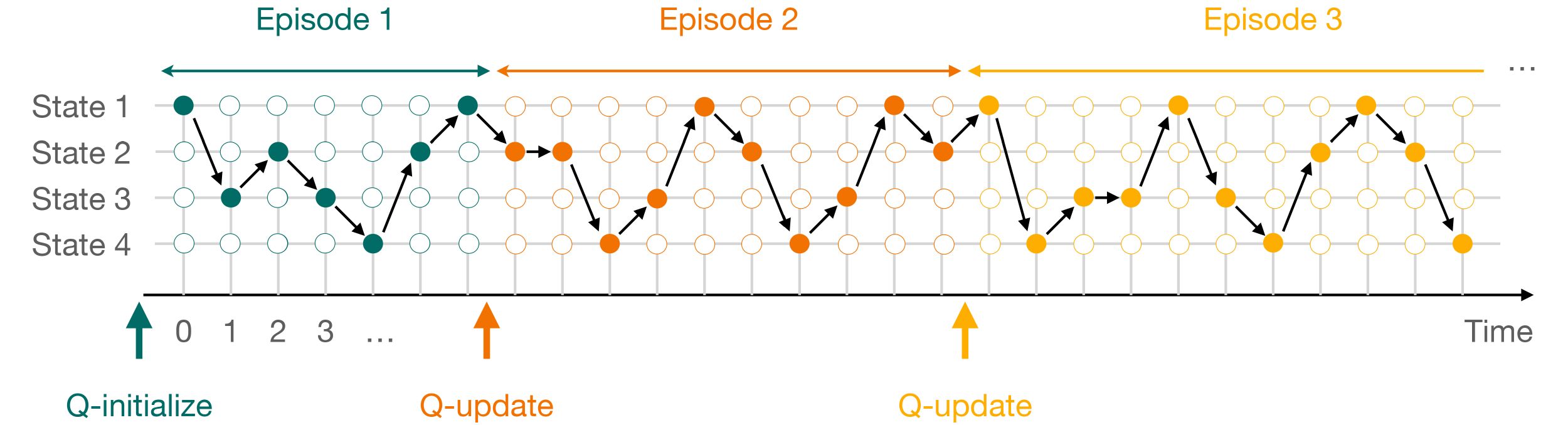}
\caption{A figurative explanation for the logit-Q learning dynamics in an SG with four states. The circles shaded represent the state visited at the corresponding stage and the agents play the associated stage game only at those times.
The arrows represent the flow of state transitions in time. The same color is used within an episode to highlight that the same normal-form games are getting played repeatedly.}
\label{fig:model}
\end{figure*}

The key property of our dynamics is to let agents \textit{play-in-episodes}. Within an episode, they play stage games whenever the associated state gets visited without any update on the Q-function estimates, as illustrated later in Figure \ref{fig:model}. Therefore, their act is always consistent with the stage game they play. We consider two update schemes for the Q-function update: In the \textit{average value update}, agents take the empirical average of the payoffs received in the repeated play of the stage games within the episode. In the \textit{frequent value update}, agents assign the value of a state as the payoff of the corresponding stage game associated with the action profile most frequently played. Both are consistent with the (desired) uncoupled nature of the dynamics. The former ensures convergence to the optimal solution approximately with an approximation error characterized. The latter ensures convergence to the optimal solution exactly. Furthermore, these update schemes make the stage games stationary within an episode and allows agents to converge to efficient equilibria at each stage game through the (independent) log-linear learning dynamics to a certain extent. Therefore, the update of continuation payoff approximates the value iteration in MDPs. Increasing episode lengths across episodes combined with the contraction property of the value iteration yield that agents can attain the globally optimum solution in the limit. The increasing episode lengths play a similar role with the step sizes vanishing at two different timescales in the two timescale learning framework used in \cite{ref:Leslie20} and \cite{ref:Sayin20}.

There are a few studies addressing efficient learning in SGs, such as \cite{ref:Wang02}, \cite{ref:Yongacoglu}, and \cite{ref:Sayin23}. In \cite{ref:Wang02}, the authors presented a learning dynamic for identical-interest SGs based on the adaptive play. Quite contrary to our uncoupled learning dynamics, theirs has a coupled structure due to the off-policy learning of game structure including a maximization over joint actions of all agents. 
Notably, in \cite{ref:Yongacoglu}, the authors focused on decentralized learning of efficient equilibrium in SGs. Their key idea is to focus on stationary pure strategies so that the underlying SG can be viewed as a (huge) normal-form game in which actions correspond to stationary pure strategies (since there are only finitely many contrary to a continuum of stationary mixed strategies). This reduces the problem to the repeated play of this (huge) normal-form game. They addressed the unknown state transitions and no access to opponent actions through learning-in-phases in which there is no update of strategies. In that sense, their dynamics have a flavor more similar with the actor-critic methods 
 where two different timescales are used in the reverse order with our approach. In other words, the approaches differ in terms of learning-in-phases vs playing-in-episodes. Lastly, in \cite{ref:Sayin23}, we address efficient learning in stochastic teams with a vanishing step size used in the value function update different from the episodic scheme presented. However, the dynamics there provably reach a neighborhood of the efficient equilibrium with a more involved convergence analysis. Recall that logit-Q learning dynamics with the frequent value update reaches the exact efficient equilibrium.

The paper is organized as follows. We describe identical-interest SGs in Section \ref{sec:model} and the new classical and independent logit-Q dynamics in Section \ref{sec:dynamics}. Then, we present the convergence results in Section \ref{sec:result}, and an overview of their proofs in Section \ref{sec:proof}. In Section \ref{sec:example}, we show the performance of the algorithm with numerical examples. We conclude the paper with some remarks in Section \ref{sec:conclusion}. An appendix includes the proof of a technical lemma.


\section{IDENTICAL INTEREST STOCHASTIC GAMES}\label{sec:model}

Formally, an $n$-agent SG is a dynamic game played over infinite horizon that can be characterized by the tuple $\langle S,A,r,p\rangle$. At each stage, the game visits a state $s$ from a \textit{finite} set $S$ and each agent $i$ simultaneously takes an action $a^i$ from a \textit{finite} set $A^i$ to collect \textit{stage-payoffs}.\footnote{The formulation can be extended to state-variant action sets rather straightforwardly.} Similar to MDPs, stage-payoffs and transition of the game across states depend only on the current state visited and current action profile $a=\{a^i\}_{i=1}^n$ played. Particularly, $p(s'|s,a)$ for each $(s,a,s')$ denotes the probability of transition from $s$ to $s'$ under action profile $a$, and $r^i:S\times A \rightarrow \mathbb{R}$ denotes the stage-payoff function of agent $i$. We specifically consider the identical-interest SGs, also known as stochastic teams, where there exists $r:S\times A \rightarrow \mathbb{R}$ such that $r^i(s,a) = r(s,a)$ for all $(i,s,a)$. 

We let agents randomize their actions according to a strategy determining the probabilities of actions to be played. They choose their strategies to maximize the expected sum of discounted stage-payoffs with the discount factor $\gamma \in [0,1)$. For example, the objective of player $i$ is given by
\be\label{eq:sum}
\Expected{\sum\nolimits_{k=0}^{\infty}\gamma^kr(s_k,a_k)},\ee
where $(s_k,a_k)$ denotes the pair of state and action profile at stage $k$ and the expectation is taken with respect to the randomness on $(s_k,a_k)$ for each $k\geq 0$. 

Notice that, in SGs, agents play a normal-form game associated with the state visited at each stage where the payoffs of these stage games, called \textit{Q-functions}, are of the form
\be \label{eq:Qdef}
Q(s,a) := \Expected{\sum\nolimits_{k=1}^{\infty}\gamma^kr(s_{k},a_{k})|s_0=s,a_0=a},
\ee
and the corresponding value of state, called value function is defined as
\be
v(s) := \Expected{\sum\nolimits_{k=1}^{\infty}\gamma^kr(s_{k},a_{k})|s_0=s},
\ee

In an SG, we say that a strategy is \textit{stationary} if it depends only on the current state (and not on the time). Similar to the existence of a stationary optimal solution in MDPs, there also exists a \textit{stationary equilibrium} in SGs such that agents do not have any incentive to change their stationary strategies unilaterally \cite{ref:Fink64}. Furthermore, specifically in identical-interest SGs, there also exists a stationary equilibrium, called \textit{efficient stationary equilibrium}, attaining the global maximum of the common objective \eqref{eq:sum}. The (unique) value function and the (unique) Q-function associated with an efficient stationary equilibrium, respectively, denoted by $v_*:S \rightarrow \mathbb{R}$ and $Q_*:S\times A\rightarrow \mathbb{R}$, satisfy the following fixed-point equations for all $(s,a)$, and $s$ :
\begin{align}
& v_*(s) = \max_{a\in A} \left\{r(s,a) + \gamma \sum\nolimits_{s'\in S}p(s'|s,a) v_*(s')\right\},\label{eq:vstar}\\
& Q_*(s,a) = r(s,a) + \gamma \sum\nolimits_{s'\in S}p(s'|s,a) \max_{a'\in A} \{Q_*(s',a')\},\label{eq:Qstar}\nn
\end{align}
and $v_*(s) = \max_{a\in A} \{Q_*(s,a)\}$ for all $s$.
The uniqueness follows from the contraction property of the right-hand side. Indeed the right-hand side in \eqref{eq:vstar} is the Bellman operator we would have if it was a (single-agent) MDP.

\section{EPISODIC (INDEPENDENT) LOGIT-Q LEARNING}\label{sec:dynamics}

An important question is \textit{how an agent would/should play}. The rich literature on learning in games has studied this question for the repeated play of the same normal-form game. 
As previously discussed, agents also play a normal-form stage game associated with the state visited at each stage in SGs. However, the payoffs of these stage games, \textit{Q-functions}, depend both on the stage-payoff received immediately, e.g.,  $\gamma^k r(s_k,\cdot)$, and the expected continuation-payoff to be received in future stages, e.g., $\gamma^k \Expected{\sum_{l=k+1}^{\infty}\gamma^{l-k} r(s_l,a_l)}$ since $s_{k+1}\sim p(\cdot|s_k,\cdot)$ as can be seen in \eqref{eq:Qdef}. The latter is ambiguous and possibly non-stationary by depending on the future play. Hence, in SGs, the agents are not necessarily playing the same stage game repeatedly whenever the associated state gets visited.

To address the ambiguity of the future play, we use the standard approach to let agents estimate the value of a state, called \textit{value function}, based on the history of their interactions as if the agents would play in the future as they played in the past. To mitigate the non-stationarity issue, we propose to let agents update their value function estimates (and therefore, the payoff of the stage games) \textit{only} at certain stages. We call the time between two consecutive updates as an \textit{episode}. Correspondingly, the agents play stationary stage games, i.e., follow the (independent) log-linear learning dynamics, described precisely later, within each episode, as illustrated in Figure \ref{fig:model}. We re-emphasize that this play-in-episodes scheme has a similar flavor with the two-timescale learning framework studied in \cite{ref:Leslie20,ref:Sayin20} in addressing the non-stationarity issue. Furthermore, it is consistent with the differences in the evolution pace of choices and preferences as studied in the evolutionary game theory literature. 

We denote the value function and Q-function estimates for episode $t=1,2,\ldots$ by $v_{(t)}:S\rightarrow \mathbb{R}$ and $Q_{(t)}:S\times A \rightarrow\mathbb{R}$, respectively.  Based on $v_{(t)}(\cdot)$, the agents can construct the Q-function estimate (and therefore, the payoff of the stage games) for the next episode $(t+1)$ according to
\be\label{eq:Q}
Q_{(t+1)}(s,a) = r(s,a) + \gamma \sum\nolimits_{s'\in S} p(s'|s,a) v_{(t)}(s'),
\ee
with $Q_{(1)}(s,a) = r(s,a)$, for all $(s,a)$.
For the update of $v_{(t)}$, we consider two schemes. Firstly, the agents can take the empirical average of payoffs received in the associated stage games, i.e.,
\be\label{eq:vave}
\ave{v}_{(t)}(s) = \sum\nolimits_{a \in A}\eta_{(t)}(s,a)Q_{(t)}(s,a),\quad \forall s,
\ee
where $\eta_{(t)}(s,a):= c_{(t)}(s,a)/c_{(t)}(s)$ is the sampled frequency of $(s,a)$, corresponding to the number of times that $a$ gets played in state $s$, $c_{(t)}(s,a)$, divided by the number of times that $s$ gets visited, $c_{(t)}(s)$, within episode $t$. Alternatively, they can estimate it as the payoff associated with the action profile played the most frequently within the latest episode, i.e., 
\be\label{eq:vfreq}
\freq{v}_{(t)}(s) = \frac{1}{|\freq{A}_{(t)}(s)|} \sum\nolimits_{a\in \freq{A}_{(t)}(s)}Q_{(t)}(s,a),\quad\forall s,
\ee
where $\freq{A}_{(t)}(s) := \argmax_{a\in A} \{\eta_{(t)}(s,a)\}$.\footnote{For a finite set $A$, we denote its number of elements by $|A|$.} Both schemes \eqref{eq:vave} and \eqref{eq:vfreq} result in a value function estimate common among agents, and therefore, the stage games always have the identical interest structure throughout the underlying SG. Furthermore, both are also consistent with the (desired) uncoupled nature of the dynamics.

In the first and the most standard version of the log-linear learning we consider, which was originally introduced by \cite{ref:Blume93}, agents respond to the latest action of others in turn. At the first play of each stage game specific to a state, the agents take arbitrary actions. Then, they keep track of the latest action played by each agent $i$ at each state $s$ until and excluding stage $k$, denoted by  $\alpha^i_k(s)$. Every agent except a randomly chosen one according to the uniform distribution over agents, say $i$, takes the latest action they have taken in the associated state. Then, agent $i$ takes action $a^i$ drawn according to the distribution $\sigma^i(Q_{(t)}^i(s_k,\cdot, \alpha^{-i}_k(s_k)))$, where $\sigma^i:\mathbb{R}^{|A^i|}\rightarrow \Delta(A^i)$ is the soft-max function given by
\be
\sigma^i(z)[a] = \frac{e^{z[a]/\tau}}{\sum_{a'} e^{z[a']/\tau}}\quad \forall z\in \mathbb{R}^{|A^i|},
\ee
where $\tau>0$ is a temperature parameter ensuring that each action gets played with some positive probability.

\begin{algorithm}[t!]
\centering
\caption{Episodic Logit-Q Dynamics} 
\label{tab:algo}
\hrule
\begin{algorithmic}[1]
\Require each agent keeps track of $\{\alpha^j_k(s)\}_{(j,s)\in [n]\times S}$ 
\For{each episode $t=1,2,\ldots$}
\State{\bfseries input:} $Q_{(t)}(s,a)$, described in \eqref{eq:Q} using $\ave{v}$ or $\freq{v}$
\For{each stage $l=\kappa_{(t)},\ldots,\kappa_{(t+1)}-1$}
\State {\bfseries input:} Current state $s_l$
\State each agent $i$ {\bfseries plays}
\[
\left\{\begin{array}{ll}
a^i(s_l) \sim \sigma^i(Q_{(t)}(s_l,\cdot, \alpha^{-i}_l(s_l))) & \mbox{if he/she is picked}\\
a^i(s_l) = \alpha^i_l(s_l)&\mbox{otherwise}
\end{array}\right.
\]
\State {\bfseries output}: Count of realizations: $c_{(t)}(s)$, $c_{(t)}(s,a)$
\EndFor
\State {\bfseries compute}: $\eta_{(t)}(s,a) = c_{(t)}(s,a)/c_{(t)}(s) \quad\forall (s,a)$, 
\State {\bfseries output}: 
\[
\left\{\begin{array}{ll}
\ave{v}_{(t)}(s) = \sum_{a}\eta_{(t)}(s,a)\ave{Q}_{(t)}(s,a) & \\
\mbox{or}\\
\freq{v}_{(t)}(s) = \frac{1}{|A_{(t)}(s)|}\sum_{a\in A_{(t)}(s)}\freq{Q}_{(t)}(s,a) &\\
\end{array} \right.
\]
\EndFor
\end{algorithmic}
\hrule
\end{algorithm} 

Though the \textit{update-in-turn} plays an important role in the convergence of the log-linear learning dynamics, certain structured relaxations minimizing the coordination among agents are also possible. One of the most critical structural relaxations is to let agents independently decide to change their actions. Therefore, in the second version of the log-linear learning we consider, agents independently respond to the latest action of others, see \cite[Section 4]{ref:Marden12} for an extended introduction. As before, at the first play of each stage game specific to a state, the agents take arbitrary actions. Then starting from the first play, they again keep track of the latest action played by each agent $i$ at each state $s$ until and excluding stage $k$, denoted by  $\alpha^i_k(s)$. Then, at each play of the stage game, in contrast to the classical log-linear learning, every agent \textit{independently} decides on whether to change their actions with some probability $\omega$. An agent, say $i$, that decides to change their action, take action $a^i$ drawn according to the distribution $\sigma^i(Q_{(t)}^i(s_k,\cdot,\alpha^{-i}_k(s_k)))$. We refer to this dynamics as \textit{independent} logit-Q dynamics.

We emphasize that our approach differs from pure algorithmic solutions used in distributed control or dynamic programming applications as it is based on a classical behavioral model. Furthermore, we emphasize that both Algorithms \ref{tab:algo} and \ref{tab:algo_ind} have an uncoupled  nature as agents do not operate on the joint action profile contrary to the coupled learning algorithms that solve the Markov team problem involving maximization over joint action profiles, e.g., Team Q-learning introduced in \cite{ref:Littman01}.

Algorithm \ref{tab:algo} and Algorithm \ref{tab:algo_ind} provide a description of the classical and independent logit-Q dynamics respectively. There, the agents know the stage-payoff function and transition probabilities but the extension into the case of unknown stage-payoffs and transition probabilities can be found in the preliminary version of this paper \cite{ref:Sayin22a}. 

\begin{algorithm}[t!]
\centering
\caption{Episodic Independent Logit-Q Dynamics} 
\label{tab:algo_ind}
\hrule
\begin{algorithmic}[1]
\Require each agent keeps track of $\{\alpha^j_k(s)\}_{(j,s)\in [n]\times S}$ 
\For{each episode $t=1,2,\ldots$}
\State{\bfseries input:} $Q_{(t)}(s,a)$, described in \eqref{eq:Q} using $\ave{v}$ or $\freq{v}$
\For{each stage $l=\kappa_{(t)},\ldots,\kappa_{(t+1)}-1$}
\State {\bfseries input:} Current state $s_l$
\State each agent $i$ independently {\bfseries plays}
\[
\left\{\begin{array}{ll}
a^i(s_l) \sim \sigma^i(Q_{(t)}(s_l,\cdot,\alpha^{-i}_l(s_l))) & \mbox{w.  p. $\omega$}\\
a^i(s_l) = \alpha^i_l(s_l)&\mbox{w. p. (1 - $\omega$)}
\end{array}\right.
\]
\State {\bfseries output}: Count of realizations: $c_{(t)}(s)$, $c_{(t)}(s,a)$
\EndFor
\State {\bfseries compute}: $\eta_{(t)}(s,a) = c_{(t)}(s,a)/c_{(t)}(s) \quad\forall (s,a)$, 
\State {\bfseries output}: 
\[
\left\{\begin{array}{ll}
\ave{v}_{(t)}(s) = \sum_{a}\eta_{(t)}(s,a)\ave{Q}_{(t)}(s,a) & \\
\mbox{or}\\
\freq{v}_{(t)}(s) = \frac{1}{|A_{(t)}(s)|}\sum_{a\in A_{(t)}(s)}\freq{Q}_{(t)}(s,a) &\\
\end{array} \right.
\]
\EndFor
\end{algorithmic}
\hrule
\end{algorithm}
\section{MAIN RESULTS}\label{sec:result}

In this section, we characterize the convergence properties of Algorithms \ref{tab:algo} and \ref{tab:algo_ind}. We say that an SG is irreducible if $p(s'\mid s,a)>0$ for all $(s,a,s')$ as in \cite{ref:Leslie20}. The following theorem characterizes the convergence of Q-function estimates in irreducible identical-interest SGs for Algorithm \ref{tab:algo}. Note that the irreducibility assumption can be relaxed by focusing on the recurrent class that the SG reaches eventually, see the preliminary version of this paper \cite{ref:Sayin22a} for detailed discussion.

\begin{theorem}{}\label{thm:asymptotic}
Given an irreducible identical-interest SG, consider that agents follow Algorithm \ref{tab:algo}. Suppose that the episode lengths go to infinity, i.e., $L_{(t)} := \kappa_{(t+1)}-\kappa_{(t)}\rightarrow \infty$ as $t\rightarrow\infty$, e.g., $L_{(t)} \propto t^2$. 
\begin{itemize}
\item[$(i)$]{If the agents follow the average value update scheme, then we have}
\be\label{eq:resultave}
 \limsup_{t\rightarrow\infty} \left| Q_{(t)}(s,a) - Q_*(s,a) \right| \leq \tau\frac{\log|A|}{1-\gamma} \quad\forall (s,a) \nn
\ee
with probability $1$.
\item[$(ii)$]{If the agents follow the frequent value update scheme, then we have}
\be\label{eq:resultfreq}
\limsup_{t\rightarrow\infty} \left| Q_{(t)}(s,a) - Q_*(s,a) \right| = 0,\quad\forall (s,a) \nn
\ee
with probability $1$.
\end{itemize}
\end{theorem}

Correspondingly, the following theorem is the counterpart of Theorem \ref{thm:asymptotic} for Algorithm \ref{tab:algo_ind}.

\begin{theorem}{}\label{thm:asymptotic_ind}
Given an irreducible identical-interest SG, consider that agents follow Algorithm \ref{tab:algo_ind}. Suppose that the episode lengths go to infinity, i.e., $L_{(t)} := \kappa_{(t+1)}-\kappa_{(t)}\rightarrow \infty$ as $t\rightarrow\infty$, e.g., $L_{(t)} \propto t^2$. 
\begin{itemize}
\item[$(i)$]{If the agents follow the average value update scheme, then we have}
\begin{align}\label{eq:resultave_ind}
 \limsup_{t\rightarrow\infty} \left| Q_{(t)}(s,a) - Q_*(s,a) \right| \leq \frac{\tau \log|A| + H(\omega)}{1-\gamma}  \nn
\end{align}
$\forall (s,a)$, with probability $1$, where $H(\omega)\in\mathcal{O}(\omega^{n})$ is characterized in \eqref{eq:H_def} and it is induced by the differences between the stationary distributions of classical and independent log-linear updates. Note that $H(\omega)$  vanishes as $\omega\rightarrow 0$.
\item[$(ii)$]{If the agents follow the frequent value update scheme, then we have}
\be\label{eq:resultfreq_ind}
\limsup_{t\rightarrow\infty} \left| Q_{(t)}(s,a) - Q_*(s,a) \right| \leq \frac{\tau M(\omega)}{1-\gamma}, \nn
\ee
$\forall (s,a)$, with probability $1$, where $ M(\omega)\in\mathcal{O}(n\log(\omega))$ is characterized in \eqref{eq:M_def} and it is induced by the differences in the maximizers of the stationary distributions of classical and independent log-linear updates. Note that $M(\omega)$ vanishes as $\omega \rightarrow 0$.
\end{itemize}
\end{theorem}
Both Theorems \ref{thm:asymptotic} and \ref{thm:asymptotic_ind} show the almost sure convergence of the Q-function estimates to a small neighborhood around $Q_*$ in both average and frequent value update schemes. 

\section{PROOFS OF THEOREMS \ref{thm:asymptotic} \& \ref{thm:asymptotic_ind}}\label{sec:proof}

The proofs are built upon the observation that if value function estimates $v_{(t)}(s)$ for each $s$ track the maximum of the Q-function estimates over action profiles, i.e., $\max_{a\in A} Q_{(t)}(s,a)$, then the evolution of $v_{(t)}$ across episodes $t=1,2,\ldots$ approximates the value iteration for MDPs (with single agent). Hence, the convergence to the global optimum follows. Therefore, we define a tracking error term, $e_{(t)}(s) := v_{(t)}(s) - \max_{a\in A} \left\{Q_{(t)}(s,a)\right\}$, on the deviation between the value function estimates and the maximum of the Q-function estimates. Then, the deviation between the optimal Q-function and the Q-function estimates, $\delQ_{(t)}(s,a) = Q_{(t)}(s,a) - Q_*(s,a)$ can be written as
\begin{align}
\delQ_{(t+1)}(s,a) =& \gamma\sum_{s'\in S} p(s'|s,a)\Big(\max_{a'\in A} \{Q_{(t)}(s',a')\} \nonumber\\
&-\max_{a'\in A} \{Q_*(s',a')\} + e_{(t)}(s') \Big).
\end{align}
Based on the non-expansiveness of the maximum function, we have
\be
\odelQ_{(t+1)} \leq \gamma \odelQ_{(t)}  + \gamma \bar{e}_{(t)},
\ee
where we define the bounds $\odelQ_{(t)} := \max_{(s,a)}|\delQ_{(t)}(s,a)|$, and $\bar{e}_{(t)} := \max_{s}|e_{(t)}(s)|$. Then, by induction, we have
\be\label{eq:induct}
\odelQ_{(T+1)} \leq \gamma^{T-t} \odelQ_{(t)}  + \sum\nolimits_{i=t}^{T}\gamma^{i-t+1} \bar{e}_{(t)} \quad\forall T\geq t \geq 0
\ee
Suppose that $\limsup_{t \rightarrow \infty}\bar{e}_{(t)} \leq c $ then for any $\epsilon >0$, there exists an episode $N$ such that
\be
\bar{e}_{(t)} < c + \epsilon \quad\forall t>N.
\ee
Now note that the first term on the right-hand side of \eqref{eq:induct} vanishes as $T\rightarrow \infty$ and the second term can be bounded by
\begin{align}
& \sum\nolimits_{i=t}^{T} \gamma^{i-t+1} \bar{e}_{(t)} \leq \frac{c+\epsilon}{1-\gamma} \quad\mbox{for any }\epsilon>0 \mbox{ and } T \geq N \nonumber.
\end{align}
Hence, we obtain 
\be
\limsup_{t\rightarrow \infty}\odelQ_{(t)} \leq \frac{c}{1-\gamma}.
\ee
Next, our goal is to characterize $c\geq \limsup_{t \rightarrow \infty}\bar{e}_{(t)}$ for Algorithm \ref{tab:algo} and Algorithm \ref{tab:algo_ind} with both average and frequent value update schemes. To do so, we will use the fact that the action profiles played at every visit to a specific state forms an irreducible and aperiodic Markov chain.
Particularly, the ergodic theorem for Markov chains, e.g., see \cite[Theorem C.1]{ref:Levin17}, yields that sampled frequency of action profiles played at the specific state $s$ converges to the stationary distribution of the Markov chain formed by them, i.e., 
\be\label{eq:etamu}
\eta_{(t)}(s,a) \rightarrow \mu_{(t)}(s,a),\quad\forall a
\ee 
with probability $1$ as the length of the episode $L_{(t)} = \kappa_{(t+1)}-\kappa_{(t)}$ goes to infinity, where $\mu_{(t)}(s,a)$ is the well known stationary distribution of the log-linear learning corresponding to $\mu_{(t)}(s,a) = \sigma(Q_{(t)}(s, a))$. We now focus on two different update schemes of Algorithm \ref{tab:algo}:

\begin{itemize}
\item[$(i)$]{In the average value update scheme,} the almost sure convergence of $\eta_{(t)}(s,a)$ to $\mu_{(t)}(s,a)$ yields that  
\be\nn
\ave{v}_{(t)}(s) \rightarrow \mathrm{E}_{a\sim \mu_{(t)}}\left[Q_{(t)}(s,a)\right]
\ee
if $L_{(t)}\rightarrow \infty$ as $t \rightarrow \infty$. This suggests that $c = \tau \log|A|$ since the stationary distribution $\mu_{(t)}(s,a)$ satisfies
\begin{equation*}
\mu_{(t)}(s) = \argmax_{\mu\in \Delta(A)} \mathrm{E}_{a\sim\mu}\left[Q_{(t)}(s,a) - \tau \log(\mu(a))\right]
\end{equation*}
and the entropy term $0\leq \mathrm{E}_{a\sim\mu}[-\log(\mu(a))] \leq \log|A|$ for any $\mu \in \Delta(A)$.
\item[$(ii)$]{In the frequent value update scheme,}
combining the definition of $\freq{A}_{(t)}(s)$ and the almost sure convergence of $\eta_{(t)}(s,a)$ to $\mu_{(t)}(s,a)$, we have, as $L_{(t)}\rightarrow \infty$,
\begin{align}
\freq{A}_{(t)}(s) &= \argmax_{a\in A} \{\mu_{(t)}(s,a)\} = \argmax_{a\in A} \{Q_{(t)}(s,a)\}. \nn
\end{align}
since there are only finitely many action profiles, i.e., $|A|<\infty$ which yields that $\mu_{(n)}(s,a)$ takes distinct values for each $a\in A$. Hence, there is a gap between the highest and the second highest values it can take. This yields that $c=0$ which completes the proof for Algorithm \ref{tab:algo}.
\end{itemize}

Next, we characterize  $c\geq \limsup_{t \rightarrow \infty}\bar{e}_{(t)}$ for Algorithm \ref{tab:algo_ind}. Almost sure convergence of $\eta_{(t)}(s,a)$ to the stationary distribution still holds. However, there is no analytical form representation of the stationary distribution, $\hat{\mu}_{(t)}(s,a)$, in the independent log-linear update. The authors in \cite{ref:Marden12} argues that the exact characterization of the stationary distribution of the Markov chain formed by the action profiles in the independent log-linear learning is quite challenging. Instead, we characterize the stationary distribution in the independent log-linear learning compared to the stationary distribution in the classical log-linear learning, i.e., $\hat{\mu}$ vs $\mu$, based on $\omega$. Given that the transition probabilities converge to the unique stationary distributions of both of the Markov chains due to their aperiodicity and irreducibility, we have
\be\label{eq:stat_lim}
\|\hat{\mu}_{(t)}(s, \cdot) - \mu_{(t)}(s, \cdot)\|_{1} = \lim_{k\rightarrow \infty}\|\hat{P}_{(t)}^k(s, \cdot) - P_{(t)}^k(s, \cdot)\|_1.
\ee
Furthermore, we have
\begin{align}
\|\hat{P}_{(t)}^k(s, \cdot) - P_{(t)}^k(s, \cdot)\|_1 \nn
&\stackrel{(a)}{=} 2 \|\hat{P}_{(t)}^k(s, \cdot) - P_{(t)}^t(s, \cdot)\|_{TV}\nn\\
&\stackrel{(b)}{\leq} 2 \Pr( (s, \hat{a}_k)  \neq (s, a_k)),\label{eq:neq}
\end{align}
where $(a)$ follows from the definition of the total variation distance, e.g., see \cite[Proposition 4.2]{ref:Levin17}, and $(b)$ follows from the coupling lemma, e.g., see \cite[Proposition 4.7]{ref:Levin17}. In the following lemma, we find a bound on \eqref{eq:neq}. Its proof follows from the proof of Lemma 2 in \cite{ref:Sayin23} with certain modifications. We provide the entire proof in Appendix \ref{app:coupling} for complete representation.

\begin{lemma}{Coupling}\label{lem:coupling}
Consider two chains $\{a_k\}$ and $\{\hat{a}_k\}$ over the same state space $A$. Both form homogenous, irreducible, and aperidic Markov chains with transition kernels $P$ and $\hat{P}$, respectively. Suppose that:
\begin{itemize}
\item[$(i)$] {Small Perturbation:} There exists some $\lambda \in (0,1]$ such that
\[
\TV{\hat{P}(\cdot\mid \hat{a}) - P(\cdot\mid  \hat{a})} \leq 1-\lambda\quad\forall  \hat{a} \in A.
\] 
\item[$(ii)$] {Matching Probability:} There exists some $\varepsilon>0$ such that
\[
\Pr(a_{k}=\what{a}_{k} \mid a_{k-1}\neq \what{a}_{k-1}) \geq \varepsilon,
\] 
meaning that these chains have a positive probability to match in a single step,  bounded from below by $\varepsilon>0$.
\end{itemize}
Then, there exists a coupling over the sequences $\{\hat{a}_k\}$ and $\{a_k\}$ such that we have
\begin{align}
\Pr\left(\hat{a}_k  \neq a_k\right) \leq (1-\varepsilon)^{k} +  \frac{1-\lambda}{\varepsilon}  \quad \forall k. \nonumber\label{eq:result1}
\end{align}
\end{lemma}

Next, we show that we can invoke Lemma \ref{lem:coupling} on the sequences of action profiles following Algorithm \ref{tab:algo_ind} denoted by $\{\hat{a}\}$ and following Algorithm \ref{tab:algo} denoted by $\{a\}$ in an episode for a fixed state s. Let $\hat{P}_{(t)}$ and $P_{(t)}$ denote the transition matrices of the Markov chains formed by the action profiles for Algorithm \ref{tab:algo_ind} and Algorithm \ref{tab:algo} respectively. Then, observe that $\hat{P}_{(t)}$ can be written as 
\be
\hat{P}_{(t)} = p_0(\omega)\mathit{I}+ p_1(\omega)P_{(t)}+ (1-p_0(\omega)-p_1(\omega))E
\ee
where $p_0(\omega) = (1-\omega)^n$ and $p_1(\omega) = n\omega(1-\omega)^{n-1}$ corresponding to the probability that zero or one of the agents choose to update their actions. In light of this observation, the transition matrix, corresponding to the case that at least one agent updates their actions, given by 
\be \label{eq:mat_id}
\tilde{P}_{(t)} = \frac{p_1(\omega)}{1-p_0(\omega)}P_{(t)}+\frac{1-p_0(\omega)-p_1(\omega)}{1-p_0(\omega)}E,
\ee
has the same stationary distribution with $\hat{P}_{(t)}$. 
$E$ is simply a perturbation matrix corresponding to more than one agents choosing to update their actions. Furthermore, Markov chain induced by $\tilde{P}_{(t)}$ is also irreducible and aperiodic. Therefore, we rather show that we can invoke Lemma \ref{lem:coupling} on the chains induced by $\tilde{P}_{(t)}$ and $P_{(t)}$. For ease of notation, we drop the $(t)$ subscript. For condition $(i)$, for any $\tilde{a} \in A$, we have
\begin{align}
\TV{\tilde{P}(\cdot\mid \tilde{a}) - P(\cdot\mid  \tilde{a})} &= \frac{1}{2}\|\tilde{P}(\cdot\mid \tilde{a}) - P(\cdot\mid  \tilde{a})\|_1\\
&= p(\omega) \frac{1}{2} \|P_{\tilde{a}} - E_{\tilde{a}} \|_1\label{eq:TV_dev},
\end{align}
where $P_{\tilde{a}}$, $E_{\tilde{a}}$ stand for the vector of the transition probabilities from $\tilde{a}$ and we define
\be
p(\omega) := \frac{1-p_0(\omega)-p_1(\omega)}{1-p_0(\omega)}.
\ee
Furthermore, we have
\be
\frac{1}{2} \|P_{\tilde{a}} - E_{\tilde{a}} \|_1 = 1 - \sum_{a\in A}\min(P_{\tilde{a}a},E_{\tilde{a}a} ),
\ee
where $P_{\tilde{a}a}$ and $E_{\tilde{a}a}$ stand for the transition probabilities from $\tilde{a}$ to $a$. For $\omega$ close to 0, $E_{\tilde{a}a}$ is always smaller than $P_{\tilde{a}a}$ and for actions $a$ that differ for more than 1 player's action from $\tilde{a}$ we have $P_{\tilde{a}a}=0$. Therefore, for $\omega$ close to 0, we have
\begin{align}\label{eq:l1_dev}
\frac{1}{2} \|P_{\tilde{a}} - E_{\tilde{a}} \|_1 &\leq 1-  \frac{(n|A^i|+1)}{1-\sum_{i=0}^1p_i(\omega)}\sum_{k=2}^n \frac{p_k(\omega)}{\binom{n}{k}} \sigma_o^k, \nonumber
\end{align}
where $p_k(\omega)=\binom{n}{k}\omega^k(1-\omega)^{n-k}$ and $\sigma_o$ is the minimum action selection probability for all $(s, a)$ which can be found by using the lower and upper bound of Q-function estimates as following
\be
\sigma_o = \frac{1}{\max_{i\in [n]}|A^i|} \exp\left(\frac{1}{\tau}(\underline{Q}-\overline{Q})\right) > 0,
\ee
where $\overline{Q}$ and $\underline{Q}$ are, respectively, upper and lower bounds on the Q-function estimates.
Then, for condition $(i)$ we define 
\begin{align}
\lambda(\omega) :=& 1- p(\omega)\left( 1-  \frac{(n|A|+1)}{1-\sum_{i=0}^1p_i(\omega)}\sum_{k=2}^n \frac{p_k(\omega)}{\binom{n}{k}} \sigma_o^k \right).\nonumber
\end{align}
Finally, for condition $(ii)$, we calculate the minimum probability, $\varepsilon$, that two chains match in one step. To do so, we focus on the actions $a$ that differ from $\tilde{a}$ for a single or none of the agent's action as only this type of transition is possible for $P$ whereas for $\tilde{P}$ any transition from an action profile to another is possible. Therefore, we can bound $\varepsilon$ by
\be
\varepsilon(\omega) \geq \frac{n|A^i|+1}{n(1-p_0(\omega))}\sum_{k=1}^n \frac{p_k(\omega)}{\binom{n}{k}} \sigma_o^{k+1}.
\ee
Therefore, we can invoke Lemma \ref{lem:coupling} as $k$ goes to infinity and combine it with \eqref{eq:stat_lim} and \eqref{eq:neq} to get
\begin{align}
\|\hat{\mu}_{(t)}(s, \cdot) - \mu_{(t)}(s, \cdot)\|_{1} &\leq \frac{2(1-\lambda(\omega))}{\varepsilon(\omega)}=:\Lambda(\omega), \label{eq:dev_result}
\end{align}
where we defined a new function of $\omega$, $\Lambda(\omega)$, for ease of notation. Notice that for $\omega$ close to $0$, $\Lambda(\omega)$ vanishes but also note that smaller $\omega$ values may slow down the convergence rate of the algorithm. Having characterized the deviation between $\hat{\mu}_{(t)}(s,a)$ and $\mu_{(t)}(s,a)$ we can now characterize $c\geq \limsup_{t \rightarrow \infty}\bar{e}_{(t)}$ for Algorithm \ref{tab:algo_ind} under two update schemes:
\begin{itemize}
\item[$(i)$]{Average Value Update Scheme:} We have an additional error term, $e_{(t)}^{\mathrm{new}}(s)$ induced by the difference between the expected value of Q-functions according to the different stationary distributions of log linear and independent log-linear updates, which can be written as 
\begin{align}
\mathrm{E}_{a\sim\hat{\mu}_{(t)}(s,\cdot)}\left[Q_{(t)}(s,a)\right]
- \mathrm{E}_{a\sim\mu_{(t)}(s,\cdot)}\left[Q_{(t)}(s,a)\right] \nonumber.
\end{align}
Using the upper bound on the Q-function estimates, $\overline{Q}$, and the result from \eqref{eq:dev_result}, we get
\be
\limsup_{t \rightarrow \infty} e_{(t)}^{\mathrm{new}}(s) \leq \Lambda(\omega)\overline{Q} =:H(\omega),\label{eq:H_def}
\ee
and correspondingly
$c = \tau \log|A| + H(\omega)$.

\begin{figure*}[t!] 
\centering
\subfigure[\textbf{Algorithm 1-ave}. Logit-Q with $\ave{v}$] {\includegraphics[width=.42\textwidth]{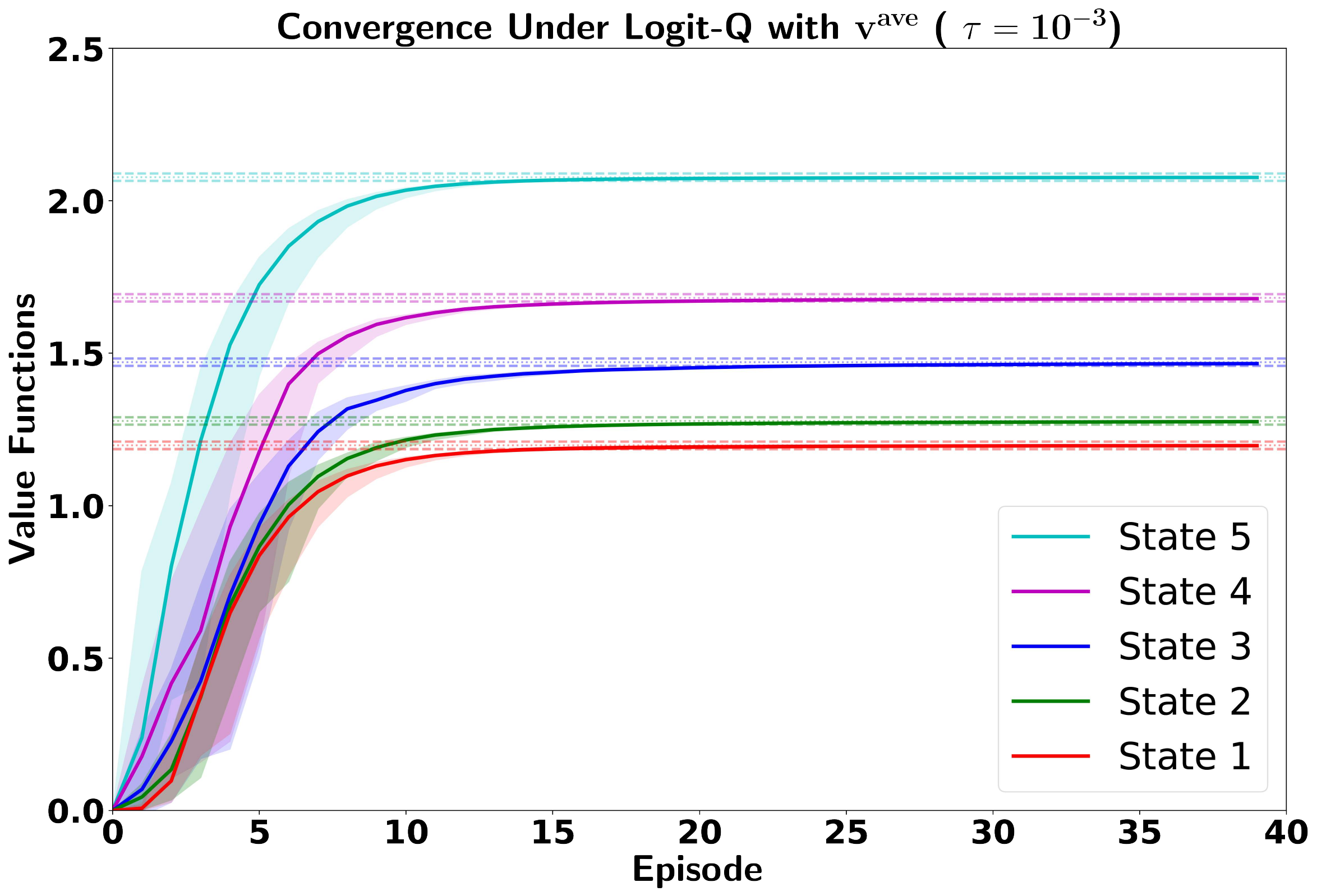}} 
\subfigure[\textbf{Algorithm 2-ave}. Independent Logit-Q with $\ave{v}$]{\includegraphics[width=.42\textwidth]{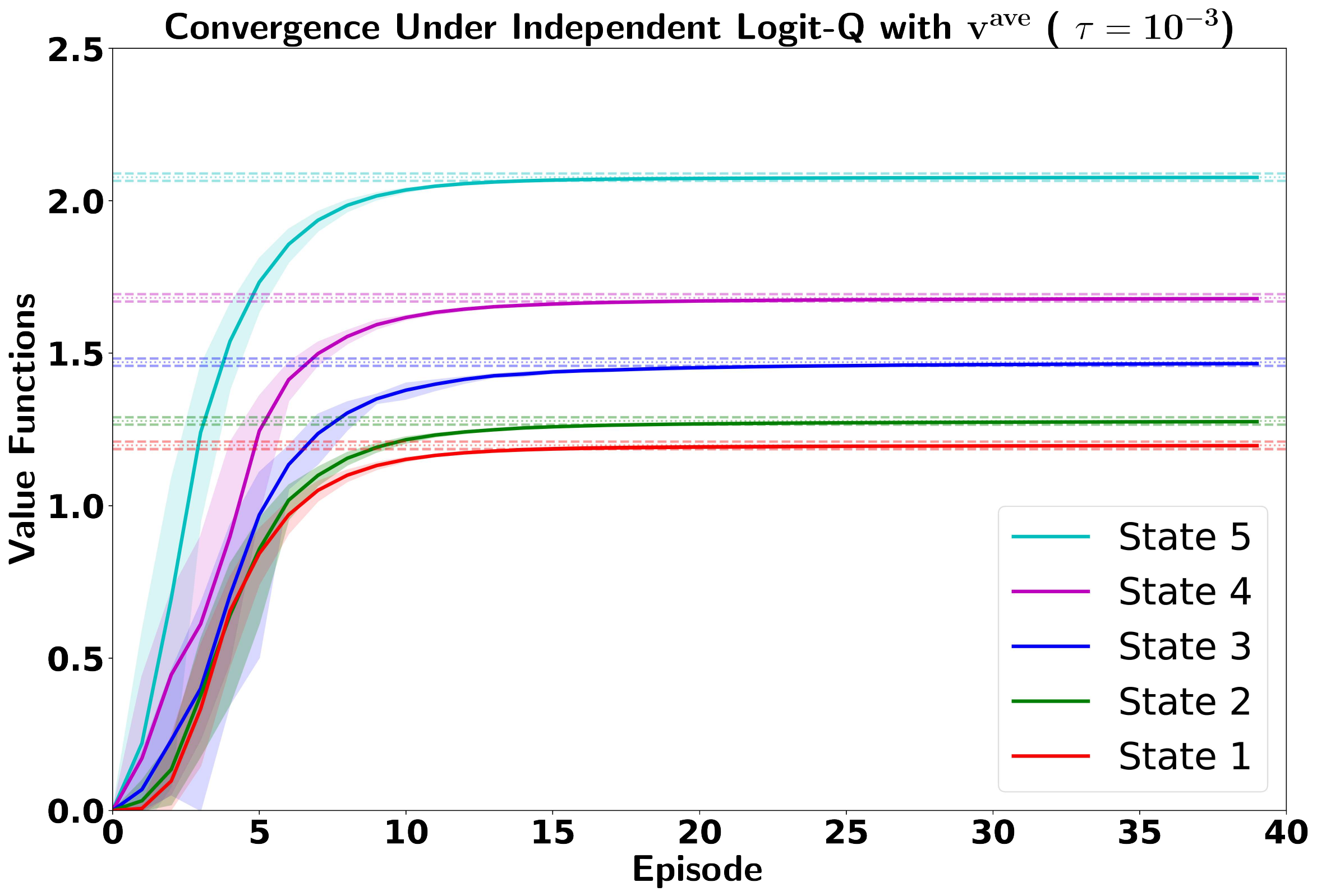}}
\subfigure[\textbf{Algorithm 1-freq}. Logit-Q with with $\freq{v}$]{\includegraphics[width=.42\textwidth]{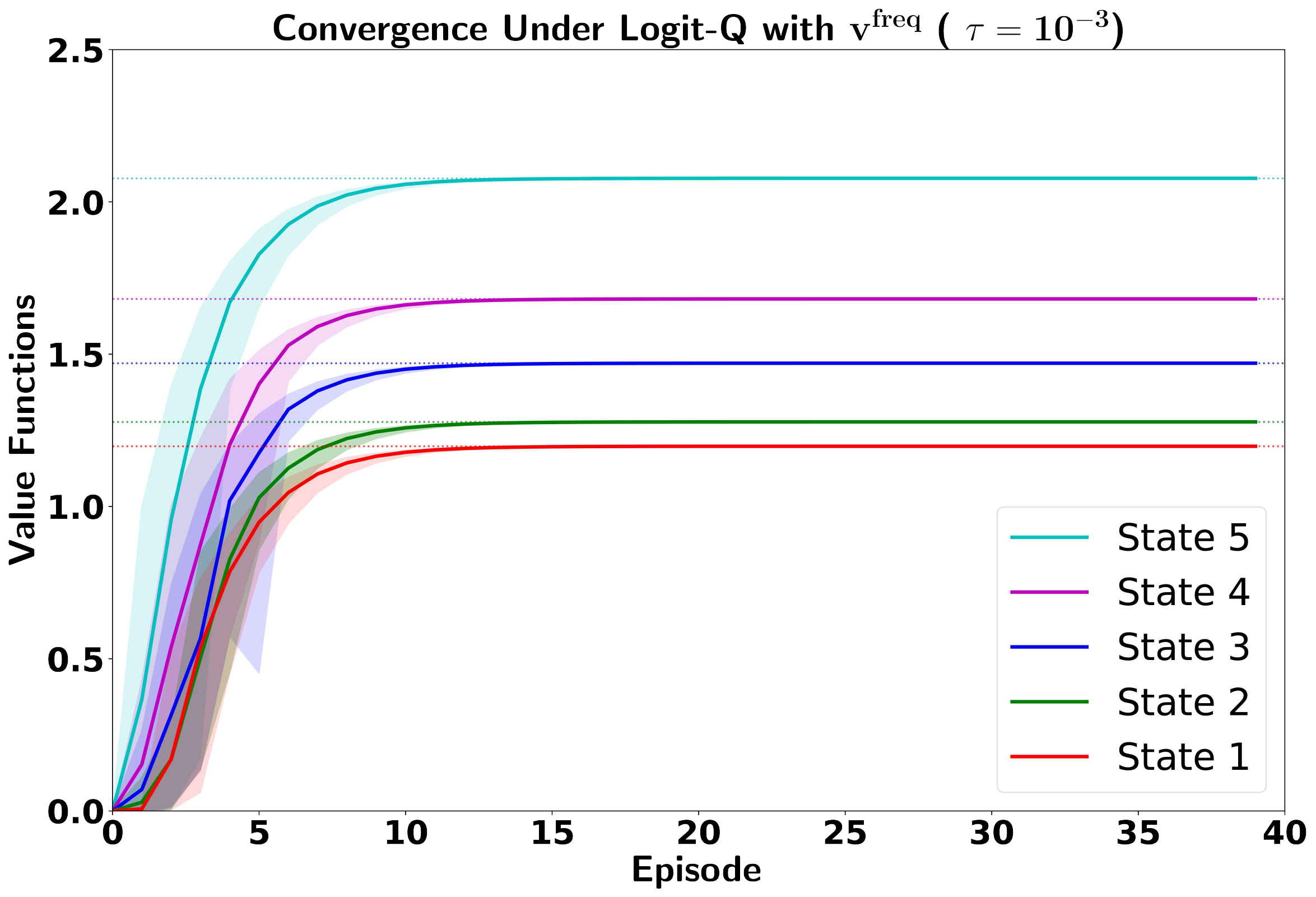}}
\subfigure[\textbf{Algorithm 2-freq}. Independent Logit-Q with $\freq{v}$]{\includegraphics[width=.42\textwidth]{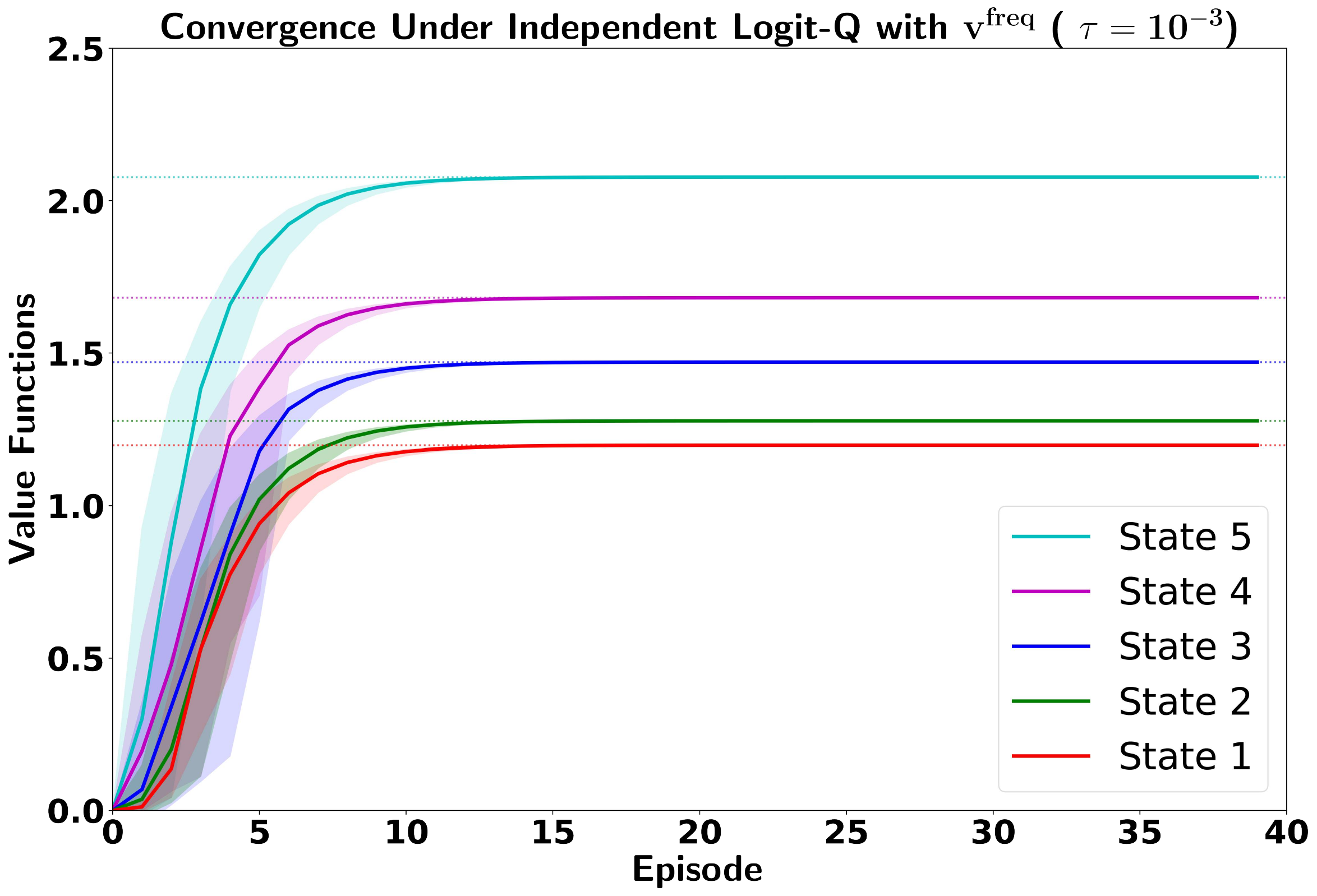}}

\caption{Convergence of the value function estimates after $20$ runs of \textbf{Algorithm 1-ave}, \textbf{Algorithm 2-ave}, \textbf{Algorithm 1-freq}, and \textbf{Algorithm 2-freq}, respectively. From bottom to top, the solid {\color{red!80!black} red}, {\color{green!80!black} green}, {\color{blue!80!black} blue}, {\color{magenta!80!black} magenta}, and {\color{cyan!80!black} cyan} curves represent the evolution of the value function estimates $v_{(t)}(s)$ for states indexed from $1$ to $5$, respectively. The error bar around the solid curves represent the maximum and the minimum values of value function estimates as a result of $20$ independent runs in \textbf{Algorithm 1}'s and \textbf{Algorithm 2}'s. The dotted lines \hdashrule[0.5ex]{0.8cm}{1pt}{1pt} represent the actual optimal state values whereas the dashed lines \hdashrule[0.5ex]{1.2cm}{1pt}{3pt}represent the bound given by the Theorem \ref{thm:asymptotic} for the update scheme with $\ave{v}$. For all of the cases, value function estimates are observed to converge to the optimal values of states.}
\vspace{-3ex}
\label{fig:simulation}
\end{figure*}
\item[$(ii)$]{Frequent Value Update Scheme:} We, again, know that $\eta_{(t)} \rightarrow \hat{\mu}_{(t)}$ as $t\rightarrow \infty$. Correspondingly, the maximizers of $\eta_{(t)}$ will converge to the maximizers of $\hat{\mu}_{(t)}$. We are interested in the Q-function values associated with the action profiles maximizing $\hat{\mu}_{(t)}$. Let $a$ be the maximizer of $\mu_{(t)}$ and $\hat{a}$ be the maximizer of $\hat{\mu}_{(t)}$. Then, we have
\be
\begin{aligned}
| \mu_{(t)}(s,a) - \hat{\mu}_{(t)}(s,\hat{a})| &\leq \max_{\tilde{a} \in A} | \mu_{(t)}(s,\tilde{a}) - \hat{\mu}_{(t)}(s, \tilde{a})| \nonumber \\
&\leq \Lambda(\omega)\nonumber
\end{aligned}
\ee
and $|\mu_{(t)}(s,\hat{a}) - \mu_{(t)}(s,\hat{a})| \leq \Lambda(\omega)$ by \eqref{eq:dev_result}. Then, by the triangle inequality, we obtain
\be
\begin{aligned}
|&\mu_{(t)}(s,a) - \mu_{(t)}(s,\hat{a})| \leq 2H(\omega). \nn 
\end{aligned}
\ee

Next, the fact that $\mu_{(t)}(s,\cdot) = \sigma(Q_{(t)}(s, \cdot))$ yields
\be
\frac{ \exp \left( \frac{1}{\tau} Q_{(t)}(s,a) \right) -\exp\left( \frac{1}{\tau} Q_{(t)}(s,\hat{a}) \right)           }{ \sum_{\tilde{a}\in A}   \exp\left( \frac{1}{\tau} Q_{(t)}(s,\tilde{a})     \right)        } \leq 2H(\omega) \nn
\ee
where we no longer need the absolute value since $a$ is the maximizer of $Q_{(t)}(s, \cdot)$. After some algebra, we obtain that $Q_{(t)}(s,a) - Q_{(t)}(s,\hat{a}) $ is upper bounded by $\tau M(\omega)$, where $M(\cdot)$ is defined by
\be
M(\omega) :=  \log \left( 2H(\omega)|A| \exp\left( \frac{(\overline{Q} -\underline{Q}) }{\tau}\right) +1 \right). \label{eq:M_def}
\ee
This completes the proof.
\end{itemize}

\section{NUMERICAL EXAMPLES}\label{sec:example}

Consider an identical-interest SG with five states and five agents, and each agent has five actions at each state. We choose the stage-payoff function and transition probabilities randomly. To this end, we first draw $r(s,a) \propto \bar{r} \cdot \mathcal{O}(s^{2})$ for each $s$, where $\bar{r}$ is chosen uniformly from $[0,1]$, and then normalize them by $\max_{(s,a)} \{r(s,a)\}$ to obtain rewards $|r(s,a)| \leq 1$ for all $(s,a)$. On the other hand, we first sample $p(s'|s,a)$ uniformly from $[0.2,1]$ and then normalize them to have $ \sum_{s'}p(s'|s,a)=1$. We set the discount factor $\gamma$, $\tau$ and $\omega$ as $0.6$, $10^{-3}$ and $10^{-2}$, respectively. 

We examine the evolution of value function estimates for both update scheme $\ave{v}$, as described in \eqref{eq:vave}, and $\freq{v}$, as described in \eqref{eq:vfreq}, under Algorithm \ref{tab:algo} and Algorithm \ref{tab:algo_ind}. We use the suffix \textbf{-ave} if agents use the average value update scheme, and \textbf{-freq} if agents use the frequent value update scheme. We set the episode lengths proportional to episode index squared, i.e., $L_{(t)} \propto t^{2}$ and note that a run for both \textbf{Algorithm 1} and \textbf{Algorithm 2} approximately takes a minute on a laptop computer equipped with a 2.6 GHz Intel i7-9750H processor, 16 GB RAM and NVIDIA GeForce RTX 2070 graphic card using Python 3.7.10 in Jupyter Notebook.

In Figure \ref{fig:simulation}, we plot the evolution of the value function estimates as a result of $20$ independent runs. 
As we have shown in Theorem \ref{thm:asymptotic} and Theorem \ref{thm:asymptotic_ind} the value function estimates converge to the optimal ones in all cases in Figure \ref{fig:simulation}. Observe that $\ave{v}$ converges to a small neighborhood around the solution, as characterized in Theorem \ref{thm:asymptotic}, while $\freq{v}$ converges to the exact solution. We deliberately plot the small neighborhood from Theorem \ref{thm:asymptotic} also for Algorithm \ref{tab:algo_ind} to show that for small values of $\omega$ Algorithm \ref{tab:algo_ind} has almost identical performance with Algorithm \ref{tab:algo}. 

\section{CONCLUSIONS}\label{sec:conclusion}

We have presented the new logit-Q and independent logit-Q learning dynamics provably attaining the social optimum in identical-interest SGs. Unlike many recent results on learning in SGs attaining possibly inefficient equilibrium, the dynamics presented converge to an efficient equilibrium almost surely, by addressing the challenge of equilibrium selection. Unlike many results on efficient learning in repeated games, our learning dynamics have convergence guarantees for SGs, by addressing the non-stationarity of the stage games induced from agents' evolving strategies. Furthermore, the dynamics presented are online in the sense that agents play and learn simultaneously in a single SG over infinite horizon without any repeated play.


\bibliographystyle{IEEEtran}
\bibliography{mybib}

\appendices

\section{PROOF OF LEMMA \ref{lem:coupling}}\label{app:coupling}
\noindent We construct a coupling over $\{\hat{a}_k\}$ and $\{a_k\}$ as follows:
\begin{itemize}[leftmargin=4mm]
\item If $\hat{a}_k \neq a_k$, then choose $\hat{a}_{k+1}$ and $a_{k+1}$ \textit{independently} according to their transition kernels $\hat{P}$ and $P$.
\item If $\hat{a}_k = a_k$, then choose $\hat{a}_{k+1}$ and $a_{k+1}$ still according to their transition kernels $\hat{P}$ and $P$, but in a \textit{coupled way} such that we have
\be\label{eq:coupled}
\Pr(\hat{a}_{k+1}\neq a_{k+1}\mid \hat{a}_{t} = a_t) = \TV{\hat{P}(\cdot\mid \hat{a}_t)-P_t(\cdot \mid a_t)}\nonumber.
\ee
Such a coupling is possible based on \cite[Proposition 4.7]{ref:Levin17}.
\end{itemize}
From condition $(ii)$, we have that
\be\label{eq:to}
\Pr(a_{k}=\what{a}_{k} \mid a_{k-1}\neq \what{a}_{k-1}) \geq \varepsilon.
\ee
Correspondingly, the probability of mismatch one-step after a mismatch is bounded from above by
\be\label{eq:upperimp}
\Pr(a_{k} \neq \what{a}_{k} \mid a_{k-1} \neq \what{a}_{k-1}) \leq 1 - \varepsilon.
\ee
Furthermore, designed coupling between these chains ensures that 
\be\label{eq:lambda}
\Pr(a_{k} = \what{a}_{k} \mid a_{k-1} = \what{a}_{k-1}) \geq \lambda \nn
\ee
based on condition $(i)$. Similarly, we can also show that the probability of mismatch one-step after a match is bounded from above by
\begin{align}
\Pr(a_{k} \neq \what{a}_{k}  \mid a_{k-1} = \what{a}_{k-1}) \leq 1 - \lambda.\label{eq:upperimp2}
\end{align} 
Next, we focus on the unconditional mismatch probability $\Pr(a_{k} \neq \what{a}_{k})$ which can be written as
\be
\begin{aligned}
&\Pr(a_{k} \neq \what{a}_{k}  \mid a_{k-1} \neq \what{a}_{k-1})\times \Pr(a_{k-1} \neq \what{a}_{k-1}) \nn\\
+ &\Pr(a_{k} \neq \what{a}_{k}  \mid a_{k-1} = \what{a}_{k-1})\times \Pr(a_{k-1} = \what{a}_{k-1}).\label{eq:induction} 
\end{aligned}
\ee
Then, \eqref{eq:upperimp} and \eqref{eq:upperimp2} yield that
\be
\Pr(a_{k} \neq \what{a}_{k}) \leq (1-\varepsilon )\Pr(a_{k-1} \neq \what{a}_{k-1}) + 1-\lambda \nn
\ee
since the fourth term on the right-hand side of \eqref{eq:induction} is bounded from above by $1$. Hence, by induction, we obtain
\begin{align}
\Pr(a_{k} \neq \what{a}_{k}) &\leq (1-\varepsilon)^k\Pr(a_{0} \neq \what{a}_{0}) + (1-\lambda) \sum_{n=0}^{k-1}(1-\varepsilon)^n\nn\\
&\leq (1-\varepsilon)^k + \frac{1-\lambda}{\varepsilon}, \nn\label{eq:ineq2}
\end{align}
where the last inequality follows by the fact that $\Pr(a_{0} \neq \what{a}_{0})\leq 1$ and the second term on the right-hand side is a partial geometric series. This completes the proof.

\end{document}